\PassOptionsToPackage{table}{xcolor}
\documentclass[conference]{IEEEtran}

\usepackage{graphicx}
\usepackage{microtype}
\usepackage{amsmath}
\usepackage{amssymb}
\usepackage{booktabs}
\usepackage{makecell}
\usepackage{array}
\usepackage{listings}
\usepackage{xcolor}
\usepackage{balance}
\usepackage{url}
\usepackage{tikz}
\usetikzlibrary{arrows.meta,positioning,calc}
\usepackage{fontawesome5}
\usepackage[hidelinks]{hyperref}

\definecolor{tblhead}{rgb}{0.859,0.902,0.961}
\definecolor{tblalt}{rgb}{0.953,0.953,0.953}
\definecolor{tbltotal}{rgb}{0.902,0.929,0.973}

\newcommand{\brk}{\allowbreak\hspace{0pt}}

\lstdefinestyle{pytool}{
  language=Python,
  basicstyle=\ttfamily\scriptsize,
  keywordstyle=\bfseries\color{blue!50!black},
  commentstyle=\itshape\color{gray},
  stringstyle=\color{green!40!black},
  showstringspaces=false,
  breaklines=true,
  breakatwhitespace=false,
  columns=fullflexible,
  frame=single,
  numbers=none,
  tabsize=2,
  captionpos=b,
  xleftmargin=2pt,
  xrightmargin=2pt,
  aboveskip=6pt,
  belowskip=6pt
}
\begin{document}

\title{Do Agent Benchmarks Do What They Say?\\
\large An Executable-Contract Audit of Tool-Using Agent Environments}

\author{
\IEEEauthorblockN{Rohith Reddy Bellibatlu, Zichong Wang, Wenbin Zhang}
\IEEEauthorblockA{Florida International University, Miami, FL, USA}
}

\maketitle

\begin{abstract}
Tool-using agents are entering settings where a wrong action carries real cost, and the benchmarks certifying them grade what each simulated tool call reports having done, assuming the tool did what its interface advertises. The audit taxonomies we survey publish no category for that assumption, and a defect beneath a score is present on every rerun. We treat a tool's advertised surfaces as an executable contract, check the implementation against it, and trace each score's provenance through the task files and evaluator code to the verdicts that derive from state a defective tool should have written. Across 34 audited mutating tools in four benchmarks we confirm seven tool defects and one evaluator property at pinned commits, one headline-eligible class per benchmark. On injected defects the checker raised no false positive in 25 flags, flagged 2 of 5 negative controls, and missed most: in 29 of 33 scored misses a clause covered the defect but no probe revealed it. The checker's own static half, run alone, flags 14 of 17 confirmed sites, several checks shaped by findings already in hand, so on those the dynamic half confirms and traces rather than discovers. On twelve further AgentDojo tools, fixed before any was audited, it found one Partial Effect the static half misses. With six held-out tools and its 7 among the 34 they complete its 25-tool mutating surface, 7 + 6 + 12,  on which at least 5 tools diverge from their advertised surface as our contracts read it, a rate for AgentDojo alone. No gold trajectory reaches either tau2-bench defect. On 1,120 paths built to isolate the telecom defect, one per task whose gold trajectory calls the tool, a number fixed by construction and not a count of affected tasks, the evaluator rewards every refuel of a suspended line and fails the repaired tool: it cannot distinguish the defect on such a path. The clearest case is a clinical benchmark whose tool tells the agent each write executed under a documented no-write design its interface does not disclose; its grader takes that message as evidence, so its action success rate records whether a request carried the expected payload, not whether any record changed.
\end{abstract}


\section{Introduction}
\label{sec:intro}

Agentic benchmarks help decide which language models ship, and the agents they power are entering settings where a mistaken action carries a real cost. A benchmark scores an agent by executing its tool calls against a simulated environment and grading what those calls report having done, from the state they left or the results they returned; either way the grade assumes each call did what its interface said, and every leaderboard position and model-selection decision built on the number inherits that assumption.\looseness=-1

That assumption is fragile: LiveClawBench names, as its motivating problem, that agent-benchmark mocks are commonly ``reduced to endpoint-level stubs that remove sessions, artifacts, state transitions, and downstream side effects'' \cite{liveclawbench26}. The benchmark audits we survey examine task artifacts, environment configuration, graders, and judges \cite{benchguard26}, \cite{aba26}, \cite{toolveritas26}, \cite{safeaudit26}.   \looseness=-1
 Of these we can check the published schemas, and none carries a category for a success signal decoupled from the state it claims to have written (27 categories: BenchGuard 14, Automated Benchmark Audit 3, Tool-Veritas 10); whether they would surface one in practice, a schema does not settle.
 
   \looseness=-1

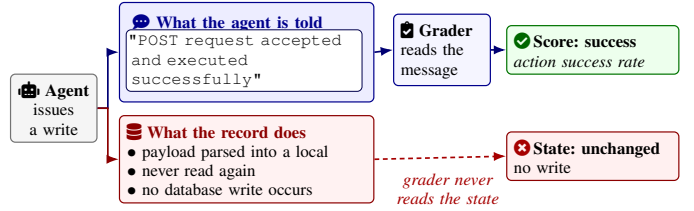
\begin{figure}[t]
\centering
\resizebox{\columnwidth}{!}{%
\begin{tikzpicture}[
  font=\scriptsize,
  panel/.style={rounded corners=2pt, align=left, inner sep=3pt, line width=0.5pt},
  told/.style={panel, draw=blue!55!black, fill=blue!7},
  does/.style={panel, draw=red!55!black, fill=red!5},
  midbox/.style={panel, draw=blue!55!black, fill=blue!4},
  okbox/.style={panel, draw=green!45!black, fill=green!7},
  nobox/.style={panel, draw=red!55!black, fill=red!6},
  actor/.style={panel, draw=black!55, fill=black!4, align=center},
  inset/.style={rounded corners=2pt, draw=blue!30!black, fill=white, line width=0.4pt,
                inner sep=2pt, font=\ttfamily\scriptsize, align=left},
  flow/.style={-{Latex[length=1.8mm]}, line width=0.7pt, draw=blue!45!black},
  dead/.style={-{Latex[length=1.8mm]}, line width=0.7pt, draw=red!65!black,
               dash pattern=on 2pt off 1.5pt},
]
\node[actor, text width=1.05cm] (agent) at (0.55,1.6) {\faRobot\ \textbf{Agent}\\ issues a write};

\node[told, text width=3.5cm, minimum height=1.45cm, anchor=north west] (told) at (1.5,3.15) {};
\node[anchor=north west, font=\scriptsize\bfseries, text=blue!55!black] at (1.58,3.1)
  {\faCommentDots\ What the agent is told};
\node[inset, text width=3.3cm, anchor=north west] at (1.6,2.75)
  {"POST request accepted\\ and executed\\ successfully"};

\node[does, text width=3.5cm, anchor=north west] (does) at (1.5,1.5)
  {{\color{red!55!black}\bfseries\faDatabase\ What the record does}\\
   $\bullet$~payload parsed into a local\\
   $\bullet$~never read again\\
   $\bullet$~no database write occurs};

\node[midbox, text width=1.2cm, anchor=west] (grader) at (5.5,2.45)
  {\faClipboardCheck\ \textbf{Grader}\\ reads the message};

\node[okbox, text width=2.3cm, anchor=west] (score) at (7.15,2.45)
  {{\color{green!40!black}\bfseries\faCheckCircle}\,\textbf{Score: success}\\ \textit{action success rate}};

\node[nobox, text width=2.3cm, anchor=west] (state) at (7.15,0.9)
  {{\color{red!65!black}\bfseries\faTimesCircle}\,\textbf{State: unchanged}\\ no write};

\draw[flow] (agent.east) -- ++(0.15,0) |- (told.west);
\draw[flow, draw=red!55!black] (agent.east) -- ++(0.15,0) |- (does.west);
\draw[flow] (told.east) -- (grader.west);
\draw[flow] (grader.east) -- (score.west);
\draw[dead] (does.east) -- (state.west);
\node[font=\scriptsize\itshape, text=red!65!black, text width=1.8cm, align=center]
  at (6.3,0.42) {grader never reads the state};
\end{tikzpicture}%
}
\caption{One MedAgentBench write: the agent is told it executed, no record changes, and the grader scores the message, not the state. The surveyed taxonomies cover the blue path and the score; the red path, what the implementation did, is what this paper tests.}
\label{fig:concept}
\end{figure}

One shipped benchmark exhibits the divergence directly. MedAgentBench \cite{medagentbench25} evaluates agents on a simulated electronic health record, and its paper documents a deliberate simplification: only GET requests reach the environment, while a POST receives a JSON-loadable sanity check after which the harness ``indicate[s] success of execution to the agent system'' (\S\,2.4, \S\,2.4.3 of that paper). Nothing the agent reads carries that disclosure: the prompt template states call syntax only, and the tool schemas are translated from FHIR, the clinical-records standard. The branch handling every write (Finding 1, Table~\ref{tab:findings}; Fig.~\ref{fig:concept}) tells the agent ``POST request accepted and executed successfully'', and no code in the repository performs the write. The defect this paper reports is not the missing write, which the maintainers chose and documented, but the agent-visible claim that it happened, which the benchmark's own write-task grader takes as its evidence.\looseness=-1

We set out to test that interface-to-implementation-to-state-transition contract directly, tool by tool, and three challenges make that hard. \textbf{No contract exists to check against}: one must be authored from the tool's advertised surfaces, and a check built on a wrong surface inherits the error. \textbf{Simplification is not defect}: simulated environments are deliberately simplified, so telling a disclosed simplification from an undisclosed one needs a principled rule, not a maintainer to ask. \textbf{A violation does not name the scores exposed to it}: that takes a trace through the benchmark's own evaluation code to the fields it reads.\looseness=-1

 \looseness=-1

That contract is also a data-quality problem: benchmark scores are a published data product the field mines and reuses, the setting in which the data-quality literature judges a product \cite{wangstrong96}, and this paper audits their provenance, in the sense dependency analysis gives the word \cite{bunemantan01}, \cite{simmhanplale05}.

The contributions are these.

(1) \textbf{Confirmed findings}: four headline-eligible (agent-visible) benchmark-class defect cells across four shipped benchmarks, from seven verified tool defects and one evaluator property, each confirmed at a pinned commit. On these findings the checker, where it runs, confirms and traces rather than discovers. A blind audit of twelve further AgentDojo tools yields one defect nobody had flagged, which the checker's static half misses, and with six held-out tools and the seven audited first completes that benchmark's 25-tool surface, on which at least 5 diverge (\S\,\ref{sec:experiments}). (2) \textbf{Six executable defect classes}, defined as checker rules rather than prose (\S\,\ref{sec:prelim}); Phantom Effect and Partial Effect have no counterpart in any surveyed audit taxonomy. (3) \textbf{An executable contract-and-checker method} (\S\,\ref{sec:method}). Its contracts are provenance-bound, each clause citing the surface it operationalizes, and are checked statically and dynamically through an adapter, or read by hand on MedAgentBench, which has none. A score-at-risk trace follows a defect to the task verdicts exposed to it, tagged by evaluator basis. Validation is pre-registered (\S\,\ref{sec:experiments}): injected defects on and off the taxonomy, negative controls with verdicts predicted before the run, gold replays, and constructed trajectories on which the shipped evaluator rewards the defect and fails its repair.\looseness=-1

\section{Related Work}
\label{sec:relwork}

 \looseness=-1

\subsection{\textbf{What a Benchmark Treats as Ground Truth}}

What settles a tool-using episode's score is either the state a call leaves behind, as in WebArena \cite{webarena24}, AppWorld \cite{appworld24}, OSWorld \cite{osworld24} and AndroidWorld \cite{androidworld25}, which grade by post-episode state rather than the agent's claim of success,   in tau2-bench \cite{tau2bench25} (gold against predicted final database state, or per-task assertions) and in AgentDojo \cite{agentdojo24} (the environment's own post-call state), or what the call returned, as in MedAgentBench \cite{medagentbench25}, which grades a write from the request the agent issued, never from a later read. Either way a score rests on the tool layer.\looseness=-1

\subsection{\textbf{Audits of Benchmarks, and the Layer Each Opens}}

Who has checked that layer? Audits divide by the layer each opens.   Task-artifact audits read instructions, gold solutions, environment configuration, grader scripts and safety coverage (BenchGuard \cite{benchguard26}; Automated Benchmark Audit \cite{aba26}; SafeAudit \cite{safeaudit26}). Verdict-layer audits ask whether a grader's decision matches the outcome: Tool-Veritas \cite{toolveritas26} on the tau2-bench family we audit, Gao and Zhou \cite{evidencebounds26} by showing a success criterion can accept a superficial proxy for the change it certifies, and JudgeSense \cite{judgesense26} under prompt rewording; Agent-Diff \cite{agentdiff26} works at that layer but grades on a state delta read from a replica API. Agent-claim audits compare what an agent reports against what was recorded (Advani \cite{falsesuccess26} against tau2-bench and AppWorld state; AgentProp-Bench \cite{agentpropbench26} by how often a corrupted parameter ends in a wrong final answer).   Practice critiques sit alongside: BetterBench \cite{betterbench24} and UTBoost \cite{utboost25} audit reported practice, Kapoor et al. \cite{kapoor24matter} without a taxonomy, the Agentic Benchmark Checklist \cite{abcchecklist25} reports $\tau$-bench \cite{taubench24} counting empty responses as success, and a construct-validity review of 445 LLM benchmarks \cite{constructvalidity25} asks whether a benchmark measures the construct it names. Each of the closest works (Table~\ref{tab:disposals}) opens a layer above the tool's state transitions, which this paper tests.\looseness=-1

   \looseness=-1

\begin{table*}[t]
\caption{The closest works on one basis: the layer each opens and, by construction of the evidence it reads, what its design takes as given. The Checklist's survey window (Jan.\ 2024--Mar.\ 2025) predates tau2-bench.}
\label{tab:disposals}
\centering
\scriptsize
\setlength{\tabcolsep}{2pt}
\begin{tabular}{@{}p{0.21\textwidth}p{0.38\textwidth}p{0.395\textwidth}@{}}
\toprule
\rowcolor{tblhead}
\textbf{Work(s)} & \textbf{Layer it opens} & \textbf{What its design takes as given} \\
\midrule
Agent-Diff \cite{agentdiff26} & State delta across a replica API, as the success criterion & The replica API that reports the delta \\
\rowcolor{tblalt}
Gao and Zhou \cite{evidencebounds26} & The grading script, asking whether an outcome is backed by stored evidence & The tool that produced the evidence \\
Tool-Veritas \cite{toolveritas26} & Grader verdict against task outcome & The state the tool wrote \\
\rowcolor{tblalt}
ToolFuzz \cite{toolfuzz25} & Tool documentation, via runtime errors and agent responses & The tool's state transitions, and any evaluator downstream \\
BenchGuard \cite{benchguard26}, Automated Benchmark Audit \cite{aba26}, SafeAudit \cite{safeaudit26} & Task artifacts, environment configuration, grader and safety coverage & No published category names the tool implementation \\
\rowcolor{tblalt}
Agentic Benchmark Checklist \cite{abcchecklist25} & Reported benchmark practice & No published category names the tool implementation \\
\bottomrule
\end{tabular}
\end{table*}

\subsection{\textbf{Contract Checking, and What Changes to Point It Here}}

The machinery for checking an implementation against a declared contract is mature: design by contract with runtime verification \cite{jcontractor05}, \cite{jass01}, pointer-state API contracts \cite{contract26}, an executable contract language \cite{icepick26}, REST invariants \cite{agora25}, frame specifications \cite{frames25}, contracts for stateful modules \cite{cogent23} and for deep-learning APIs \cite{dlcontract23}. Our Ignored Argument and Partial Effect rules are metamorphic relations in all but name \cite{segura18}, \cite{segura16}, \cite{chen18}; stateful sequence exploration is established for REST APIs \cite{restler19}, \cite{godefroid20}; PolDet \cite{gyori15} and iDFlakies \cite{idflakies19} anticipate Reset Leak through a comparable oracle.\looseness=-1

A second line applies this vocabulary to agents and places the contract on the trusted side: four works \cite{toolgate26}, \cite{abc26}, \cite{contract2tool26}, \cite{contractbench26} learn one from documentation and traces, enforce one on a tool result at runtime, or score an agent against one, none auditing the implementation behind it; the 34-fault taxonomy \cite{faulttaxonomy26} and the eight agent-code defect types of \cite{agentdefects26} likewise take the agent, not its environment, as their object.   ToolFuzz \cite{toolfuzz25} comes nearest, testing LangChain tools against their own documentation, but its taint fuzzer instruments only argument handling and its oracles are a tool runtime error or a wrong agent response: no state-transition predicate, no trace into a benchmark's evaluator. The technique transfers and the target does not: here either side, advertised text or code, may be at fault.  \looseness=-1

\section{Notations and Preliminaries}
\label{sec:prelim}

\paragraph{The tuple, and the defect taxonomy} Every defect class is defined over one or more tuples \texttt{(pre, post, args, result)}, one per probed call or call sequence from a fresh environment: the canonical state snapshots before and after a call, the call's arguments, and its return value. Table~\ref{tab:taxonomy} defines six such classes, each an executable checker rule typed at the tool boundary; the rule, not its gloss, is the definition. A contract reads success from \texttt{result} through a \emph{success signal} and may declare it \emph{biconditional}, success if and only if every advertised, non-deferred effect applied; the declaration is opt-in and justified per contract, and Phantom Effect is checked only under it. All six classes were fixed after the tau2-bench and MedAgentBench findings and before the other two benchmarks were audited, so those the findings instantiate fit them by construction, and the typing is a checking discipline, not a partition. By source reading, Finding 5 satisfies both the Ignored Argument and the Phantom Effect rule on one call, though its biconditional never fires on AgentDojo's shipped environment data (\S\,\ref{sec:experiments}). The Ignored Argument rule is deliberately state-only, since a conjunctive rule would miss AgentDojo's \texttt{reserve\_car\_rental}, which drops \texttt{end\_time} from state while interpolating it into the success string (Finding 6); a result that varies where state does not is an aggravating signal, never a class.\looseness=-1

\begin{table*}[t]
\caption{Six executable defect classes. Status: observed in a shipped benchmark, or exercised only by injected mutants.}
\label{tab:taxonomy}
\centering
\scriptsize
\begin{tabular}{@{}p{0.16\textwidth}p{0.62\textwidth}p{0.10\textwidth}@{}}
\toprule
\rowcolor{tblhead}
\textbf{Class} & \textbf{Checker rule (over \texttt{pre}, \texttt{post}, \texttt{args}, \texttt{result})} & \textbf{Status} \\
\midrule
Phantom Effect & success signal true $\land$ advertised effect delta absent & field-observed \\
\rowcolor{tblalt}
Unenforced Precondition & precondition predicate false $\land$ (no error signal $\lor$ state mutated) & field-observed \\
Ignored Argument & post-state invariant under variation of an argument advertised as effective & field-observed \\
\rowcolor{tblalt}
Partial Effect & $\geq\!1$ advertised effect predicate holds $\land$ $\geq\!1$ fails on the same call & field-observed \\
Invariant Break & environment invariant false after a legal call sequence & mutation-only \\
\rowcolor{tblalt}
Reset Leak & snapshot after reset $\neq$ initial snapshot & mutation-only \\
\bottomrule
\end{tabular}
\end{table*}

\paragraph{The benign-simplification principle} One criterion separates a maintainer's intentional simplification from an undisclosed divergence: \emph{a simplification is benign exactly when it is advertised; the defect is never the simplification but the undisclosed divergence between what the interface tells the agent and what the implementation does.} Three consequences follow. The tier decides, not the intent. tau2-bench records Finding 3's unreleased seats in a log line (\texttt{airline/tools.py:367}) and, in another tool, a deferred flight-database update in a comment (line 689). Neither reaches the agent, so neither is disclosure: Finding 3 stays a finding at the maintainer-annotated tier (\S\,\ref{sec:method}), and the line-689 case is a true negative only because no advertised surface promises the update. Either side of a divergence may be repaired, a docstring fix counting as fully as a code fix, and such a response counts as a resolved finding. And the advertised surface is the right baseline, the agent's behavior being the measured quantity: the scope is \textbf{agent-side measurement validity}, not \textbf{consumer-side label validity}, since a metric name has no fixed audience and agent-visible text is a discrete artifact.\looseness=-1

\paragraph{An evaluator-layer property outside the taxonomy} MedAgentBench's write graders exhibit what we name Ungrounded Oracle: the oracle decides a state change from the actor's claim, not the state. Observed at Finding 4, it is not a seventh class: the six are typed over \texttt{(pre, post, args, result)} at the tool boundary, oracle grounding over grader source code.\looseness=-1

\section{Method}
\label{sec:method}

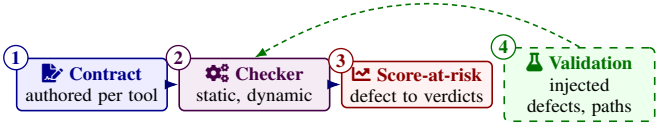
\begin{figure}[t]
\centering
\begin{tikzpicture}[
  font=\scriptsize,
  st/.style={rounded corners=2pt, align=center, inner sep=2pt,
             text width=1.85cm, minimum height=0.55cm, line width=0.6pt},
  flow/.style={-{Latex[length=1.8mm]}, line width=0.7pt, draw=blue!45!black},
  back/.style={-{Latex[length=1.6mm]}, line width=0.6pt, draw=green!45!black,
               dash pattern=on 2pt off 1.5pt},
  badge/.style={circle, fill=white, draw, line width=0.5pt, inner sep=0.3pt,
                font=\scriptsize\bfseries, minimum size=0.32cm},
]
\node[st, draw=blue!55!black, fill=blue!7] (a) at (0,0)
  {{\color{blue!55!black}\bfseries\faFileSignature\ Contract}\\ authored per tool};
\node[st, draw=violet!55!black, fill=violet!8] (b) at (2.15,0)
  {{\color{violet!55!black}\bfseries\faCogs\ Checker}\\ static, dynamic};
\node[st, draw=red!55!black, fill=red!5] (c) at (4.30,0)
  {{\color{red!55!black}\bfseries\faChartLine\ Score-at-risk}\\ defect to verdicts};
\node[st, draw=green!45!black, fill=green!7, dashed] (d) at (6.45,0)
  {{\color{green!45!black}\bfseries\faFlask\ Validation}\\ injected defects, paths};
\node[badge, draw=blue!55!black, text=blue!55!black] at (a.north west) {1};
\node[badge, draw=violet!55!black, text=violet!55!black] at (b.north west) {2};
\node[badge, draw=red!55!black, text=red!55!black] at (c.north west) {3};
\node[badge, draw=green!45!black, text=green!45!black] at (d.north west) {4};
\draw[flow] (a) -- (b);
\draw[flow] (b) -- (c);
\draw[back] (d.north) to[out=150,in=30] (b.north);
\end{tikzpicture}
\caption{Method overview: stages 1 to 3 are \S\,\ref{sec:method}'s subsections, not every benchmark takes all three (see text), and validation (\S\,\ref{sec:experiments}) measures the checker and, on constructed paths, the evaluator.}
\label{fig:pipeline}
\end{figure}

  Not every benchmark takes the three stages of Fig.~\ref{fig:pipeline}. tau2-bench and AgentDojo take all three: a contract per tool, the checker through an adapter, and the trace into their evaluators. MM-ToolSandbox takes the first two, its trace stopping after the defect-to-field step because its grading scenarios live in the \texttt{appworld} package, which this project cannot clone. MedAgentBench has neither adapter nor contract file: its clauses are read by hand from source at the pinned commit against the same predicates, and its trace runs on grader source, classifying each write grader by what it reads. Every later claim that the checker reproduced or found a defect concerns the first three benchmarks; MedAgentBench's findings are hand readings.\looseness=-1

\subsection{The Contract Specification}
\label{sec:contract}

  A contract is one YAML file per tool, authored from the tool's advertised surfaces and checked by machine. Its clauses are preconditions (\texttt{pre.}), effects (\texttt{eff.}), frame clauses (\texttt{frame.}) asserting that state outside the advertised effect is unchanged, effective-argument declarations (\texttt{arg.}), invariants and the success signal, each citing its \emph{provenance}, the advertised text it derives from, if any. The shipped \texttt{spec/\brk contracts/\brk tau2/\brk refuel\_data.yaml} at tau2-bench commit \texttt{c3398666} shows the chain. The docstring of \texttt{refuel\_data} (line 613; Table~\ref{tab:findings}, row 2) advertises ``Checks: Line status must be Active, Customer owns the line.'' The author turns the first check into the clause \texttt{pre.line\_active}, citing line 613 and the quote ``Line status must be Active'', with predicate \texttt{any(l.status == 'Active' for l in pre.lines if l.line\_id == args.line\_id)}. Its \texttt{on\_precondition\_violation} block, cited to the ``Raises'' entry at line 625 (``if checks fail''), requires a \texttt{ValueError} and no state delta. The probe generator (\S\,\ref{sec:checker}) finds, among assignments harvested from the pre-snapshot, one falsifying the predicate, customer \texttt{C1003}'s own line \texttt{L1009}, status \texttt{Suspended}; the checker calls the tool with \texttt{gb\_amount} 40.0 and observes no error, \texttt{data\_refueling\_gb} moved from 0.0 to 40.0, a new draft bill of \$80.00, and the return ``Successfully added 40.0 GB of data for line L1009 for \$80.00'': VIOLATES, Unenforced Precondition. That call, diff and result are its \emph{witness}, stored in the \emph{findings ledger} of VIOLATES rows. The enforcing branch is commented out (lines 629--630).\looseness=-1

The division of labour is fixed. Every clause is written by a person or by an annotator session, a model session (\S\,\ref{sec:q5}), reading the advertised surface; the predicate is the interpretive step, and nothing in the pipeline authors one. A drafting aid emits \emph{draft clauses}, placeholders with a \texttt{\_draft} id suffix and the constant predicate \texttt{True}, outside every count until a person writes the predicate. The validator runs eight checks per file, among them that every predicate compiles under a whitelisted grammar with no free names, that every provenance quote occurs verbatim at its cited file and line in the pinned commit (check 4), and that a cited tool-return line is a return or a raise rather than a log call or a comment, prompt templates facing a weaker file-level test (check 8).\looseness=-1

Clauses fall into three grounding tiers by provenance: \textbf{agent-visible} (docstring, schema, prompt template, tool return, README, external standard); \textbf{maintainer-annotated}, a comment, log line or TODO no agent ever sees; and \textbf{author-assigned}, a clause with no surface behind it whose semantics we assigned, as \texttt{refuel\_data}'s two frame clauses are. The schema flags author-assigned and draft clauses alike as \texttt{inferred: true}; only draft clauses carry the suffix and placeholder. A benchmark-class cell (\S\,\ref{sec:setup}) is headline-eligible only when its clause is agent-visible: for tool-return and prompt-template provenance that is check 8's outcome or, with no contract file, a hand reading at the pinned commit; the other surfaces are presumed visible rather than tested.\looseness=-1

\subsection{The Conformance Checker}
\label{sec:checker}

  The checker has a static half and a dynamic half, and only the dynamic half issues verdicts. The static half is five checks over a tool's syntax tree that execute nothing: a literal success return in a function that writes state, a commented-out guard, a parameter read only in a log call or a returned f-string, a truthiness test on a parameter whose type admits a meaningful falsy value, and a verb-paired function (book/cancel) that never references a field its partner writes. A flag names a file, function and line with a class hint, a candidate for a person to read, never a verdict. Run alone, it is Q3's \emph{scanner}.\looseness=-1

The dynamic half runs one contract through an adapter that hosts the benchmark in a subprocess and exposes four operations: build a fresh environment, snapshot its state as JSON, invoke a tool, reset. From a fresh pre-snapshot the probe generator derives the plan: per precondition one assignment satisfying it and one violating it, one happy-path assignment satisfying all, per argument declared effective a set of assignments differing only in that argument, and a joint-falsy assignment when two or more effective arguments admit a falsy value, all values harvested from the snapshot. Each probe runs in a fresh environment as snapshot, invoke, snapshot. Effect clauses are evaluated on every successful precondition-satisfying call, and frame clauses on the happy-path call, since a raise leaves \texttt{post} equal to \texttt{pre}. A violating call feeds the precondition-enforcement rule. The variant sets feed the Ignored Argument rule, which fires when the canonical post-state is identical across at least two successful calls with distinct values. Reset compares the post-\texttt{reset()} snapshot with the initial one, and invariants are evaluated after a legal call sequence.\looseness=-1

Each evaluation yields one of three verdicts per clause. CONFORMS: the predicate held. VIOLATES: it did not, recorded with a witness and the class Table~\ref{tab:taxonomy} assigns. A failing effect clause is Ignored Argument when its predicate reads an effective argument's value and Partial Effect otherwise, as is a frame violation on a call where an effect clause also fails. Phantom Effect is assigned only when a contract declares its success signal biconditional (\S\,\ref{sec:prelim}) and every testable effect clause fails on one call. UNTESTABLE: the clause could not be graded, with a reason code: \texttt{no\_observable\_state} when no probe reached the state, \texttt{predicate\_type\_error}, \texttt{no\_reset\_path} or \texttt{adapter\_unsupported}. A tool rolls up as VIOLATES if any clause violates, UNTESTABLE if no clause could be graded, and otherwise CONFORMS with its ungraded clauses still counted. UNTESTABLE stays in every clause denominator, and its share is large: 96 of 180 clause rows on tau2-bench, 11 of 60 on AgentDojo's first seven contracts, 4 of 27 on MM-ToolSandbox, every one \texttt{no\_observable\_state}. The two halves combine only in a person: a static flag or a VIOLATES row becomes a finding by source reading at the pinned commit.\looseness=-1

\subsection{Score-at-Risk as a Provenance Trace}
\label{sec:trace}

This stage treats a score as data with a lineage \cite{bunemantan01}, \cite{simmhanplale05}, the sense in which a clause's provenance is also meant. Each task verdict is a \emph{derived record}; its \emph{source records} are the task's entry in the pinned task file, the tool implementations that write state and the evaluator code; its \emph{derivation} is the set of state fields and transcript items the evaluator reads. The trace recovers derivations statically and runs backwards: given a confirmed defect, which verdicts derive from state the defective tool should have written. A script performs it with no model, over the 1,228 rows of the score-at-risk ledger, in three steps. Defect to field: a VIOLATES clause's compiled predicate, never its source string, yields the state paths the tool should have written and did not, or wrote when a precondition should have blocked the write. Field to evaluator: a read-site extractor parses the evaluation code the same way for the paths it reads, each read site classified as live simulator state or agent transcript. Evaluator to tasks: every task whose derivation includes a field in that intersection is enumerated, exhaustively. Each row's \emph{basis} tag says how its evaluator reads the field: the written attribute (\texttt{exact\_field}), its collection only (\texttt{collection\_only}), a whole-database hash (\texttt{whole\_state\_hash}), or unresolvably (\texttt{unresolved}).\looseness=-1

  Three levels are distinguished, and only these words name them. A task is \emph{exposed} when its evaluator reads a field the defect-to-field step names, resolved when the extractor reads the field at the read site or forced by the fallback below when it cannot; a write grader that reads the tool's return instead, MedAgentBench's, is classified by what it reads and gets no exposure count. A trajectory \emph{reaches} the defect when it calls the defective operation under its trigger condition, a refuel of a line that is not Active. A \emph{verdict change} is the evaluator answering differently than under a conforming tool. Table~\ref{tab:scoreatrisk} counts exposure only; gold replay measures reach on gold trajectories; the positive controls establish reach and verdict change on constructed paths; and an exposure count is called a bound only when every row in it is resolved.\looseness=-1

The extractor resolves direct attribute access on the pre- or post-state, dotted string constants naming state paths, and helpers resolvable by bare name within three hops. Exposure over-approximates verdict change only where the oracle is state-grounded and the evaluator is in that form. Any read the extractor cannot resolve, such as computed attribute access, a deeper or external helper or a whole-object comparison, forces the row in as exposed at confidence \texttt{unresolved}. This fallback was fixed before it was applied and can only widen the population. \looseness=-1

\section{Experiments}
\label{sec:experiments}

\subsection{Experimental Setup}
\label{sec:setup}

That method runs against four shipped agent benchmarks at pinned commits: MedAgentBench \cite{medagentbench25} (clinical EHR), tau2-bench \cite{tau2bench25} (airline, retail and telecom customer service), AgentDojo \cite{agentdojo24} (security-focused tool use) and MM-ToolSandbox \cite{mmtoolsandbox26} (multi-modal tool use). A tool is in scope when its advertised surface claims a state change or the benchmark marks it as mutating, enumerated by one rule per benchmark: tau2-bench's \texttt{mutates\_state} flag; on AgentDojo, which carries no tag, a read of each registered tool body for a write to injected state; and a hand read of the advertised surface elsewhere. The rules agree with the advertised criterion on every tool they admit; only MedAgentBench's write path enters on advertisement alone, its implementation changing nothing.\looseness=-1

\paragraph{The anchor audit and its coverage} The \emph{anchor audit}, 34 tools, is the audit behind the findings, Tables~\ref{tab:perbench}--\ref{tab:scoreatrisk} and the trace; tau2-bench and MedAgentBench entered it on known defects, the other two to test whether the taxonomy generalized. Its coverage differs by benchmark. It takes every in-scope tool of MedAgentBench (3) and of tau2-bench's airline, retail and telecom domains (19), three of its five. On AgentDojo it takes 7 of the 25 mutating tools the rule above enumerates, the three finding-bearing tools and four not known to be defective; the held-out and blind audits take the other 18. On MM-ToolSandbox it takes 5, the finding-bearing \texttt{venmo\_social} and four of the 13 mutating tools of its self-contained tier, and nothing here completes that surface. The eight with no finding were hand-chosen, by no prior rule, so the mutation corpus was not all defective: one per AgentDojo suite, and two reminder, one calendar and one settings tool on MM-ToolSandbox. The anchor audit samples neither surface; there its results are existence proofs, not rates.     \looseness=-1

\paragraph{Further populations} Four more are reported where used, never pooled into the anchor totals. \emph{Held-out}, 6: AgentDojo banking and Slack tools fixed by suite in a plan before any was audited, prior knowledge declared. \emph{Blind}, 12: the rest of AgentDojo's v1 mutating surface, fixed by a plan addendum before any was audited, 11 with no prior knowledge, testing discovery beyond code already suspected. \emph{AgentDojo complete surface}, 25 = 7 + 6 + 12: the population of the one per-benchmark rate. \emph{Scanner population}, 49 = 31 + 6 + 12: the 31 anchor tools with a per-tool function boundary (MedAgentBench's 3 share one dispatch method) plus held-out and blind, used only to score the static scanner.\looseness=-1

 \looseness=-1

\paragraph{Units}   Five units recur and nest. A \emph{clause} is one predicate in a contract, with id and provenance (\S\,\ref{sec:contract}). A \emph{VIOLATES row} is one clause failing on one probe, written to the findings ledger with its witness. One clause can yield several rows (\texttt{cancel\_\brk reservation}'s \texttt{eff.seats\_released} fails on two probes), one defect several clauses. A \emph{finding instance} is one defect at one site, confirmed by reading source at the pinned commit, however many rows evidence it. The ledger's 12 VIOLATES rows sit on seven tools: six confirmed instances, Findings 2, 3, 5, 6, 7 and 8, and one unadjudicated candidate, \texttt{suspend\_line}'s \texttt{arg.reason}. That clause is author-assigned: a required \texttt{reason} is logged, never persisted, and whether the docstring advertises persistence rather than a log entry is contested. MedAgentBench's two instances, Findings 1 and 4, have no rows, since it has no adapter.\looseness=-1

A \emph{benchmark-class cell} is one defect class, or the evaluator property, in one benchmark. It is the reporting unit, so that eight bugs in one copied helper count once; the eight instances populate seven cells and the candidate an eighth (Table~\ref{tab:perbench}). A \emph{source site} is one line or function the scanner can name, used only to score it. The anchor findings hold 11, six of them Finding 5's (five truthiness guards and one constant return); its held-out second instance adds four and the two blind calendar tools one each, 17 in all.\looseness=-1

\begin{table}[t]
\caption{Anchor-audit totals per benchmark; its coverage per benchmark is stated in \S\,\ref{sec:setup}.}
\label{tab:perbench}
\centering
\scriptsize
\begin{tabular}{@{}>{\raggedright\arraybackslash}p{0.285\columnwidth}>{\raggedleft\arraybackslash}p{0.13\columnwidth}>{\raggedleft\arraybackslash}p{0.12\columnwidth}>{\raggedleft\arraybackslash}p{0.12\columnwidth}>{\raggedleft\arraybackslash}p{0.28\columnwidth}@{}}
\toprule
\rowcolor{tblhead}
\textbf{Benchmark} & \textbf{Tools} & \textbf{Mut.} & \textbf{Impl.} & \textbf{Classes (HL)} \\
\midrule
MedAgentBench & 3 & 3 & 1 & 2 (1) \\
\rowcolor{tblalt}
tau2-bench & 19 & 19 & 3 & 3 (1) \\
AgentDojo & 7 & 7 & 3 & 2 (1) \\
\rowcolor{tblalt}
MM-ToolSandbox & 5 & 5 & 1 & 1 (1) \\
\midrule
\rowcolor{tbltotal}
\textbf{Total} & \textbf{34} & \textbf{34} & \textbf{8} & \textbf{8 (4)} \\
\midrule
\multicolumn{5}{@{}p{\columnwidth}@{}}{Tools: audited; Mut.: in scope as mutating; Impl.: distinct implementations behind the cells; Classes: every cell, candidate included; (HL): headline-eligible.} \\
\bottomrule
\end{tabular}
\end{table}

  Every experiment below was registered in a plan committed before it produced any number: the mutation arms, replay, positive and negative controls, scanner population and scoring rule, held-out and blind populations and trace fallback. Analyses added after a result, such as the miss diagnostic and the comparability filter, are labelled post hoc where they appear. The plans, freeze tags and commits are listed in the artifact, archived at Zenodo, DOI \href{https://doi.org/10.5281/zenodo.22182792}{10.5281/zenodo.22182792}, with development history at \url{https://github.com/rohithreddybc/tool-contract-conformance} (MIT license); audited benchmarks are pinned by upstream commit, not redistributed. \texttt{make reproduce-results} re-derives every table and audits every numeric claim from a bare checkout, offline, with no API keys or model calls, from the committed score-at-risk ledger; \texttt{make tables-gated} re-derives that ledger from the pinned clones and the SHA-256-pinned \texttt{refsol.py}.\looseness=-1

\subsection{Main Results}

  Five questions, one block each, in this order: what the audit finds (Q1); which verdicts are exposed to what it finds, and whether any changes (Q2); what the checker adds beyond reading source (Q3); how reliable the checker is (Q4); whether contract authoring is stable (Q5).\looseness=-1

\subsubsection{Q1, what the audit finds} Eight confirmed instances in seven cells, plus the unadjudicated \texttt{suspend\_line} candidate as an eighth (Table~\ref{tab:findings}). The dynamic checker reproduces three of the four headline cells as VIOLATES (tau2-bench's Unenforced Precondition, AgentDojo's and MM-ToolSandbox's Ignored Argument), MedAgentBench's being the hand reading. Four cells are reported but not headline, each for one reason: MedAgentBench's Ungrounded Oracle is an evaluator property, tau2-bench's Ignored Argument is the candidate and its Partial Effect maintainer-annotated, and AgentDojo's Phantom Effect is source-observed only.\looseness=-1

\begin{table*}[t]
\caption{Confirmed findings (instance level); each row is checkable at its pinned commit from the quoted string.}
\label{tab:findings}
\centering
\scriptsize
\setlength{\tabcolsep}{2pt}
\begin{tabular}{@{}>{\centering\arraybackslash}p{0.012\textwidth}p{0.105\textwidth}p{0.2\textwidth}p{0.115\textwidth}p{0.155\textwidth}p{0.375\textwidth}@{}}
\toprule
\rowcolor{tblhead}
\textbf{\#} & \textbf{Bench. @ commit} & \textbf{Tool : line} & \textbf{Class} & \textbf{Tier} & \textbf{Quoted evidence and disposition} \\
\midrule
1 & MedAgentBench @ \texttt{9926011} & POST branch, \texttt{\_\_init\_\_.\brk py:85-91} & Phantom Effect & agent-visible (\texttt{prompt\_\brk template}, \texttt{tool\_return}) & ``\ldots executed successfully'' returned with no write; payload never re-read. \\
\rowcolor{tblalt}
2 & tau2-bench @ \texttt{c3398666} & \texttt{refuel\_data}, \texttt{telecom\brk /tools.py:\brk 607-657} & Unenforced Precondition & agent-visible (\texttt{docstring}) & ``must be Active'' check commented out (selective disable). \\
3 & tau2-bench @ \texttt{c3398666} & \texttt{cancel\_\brk reservation}, \texttt{airline\brk /tools.py:\brk 315, 363-368} & Partial Effect & \textbf{maintainer-annotated: not headline-eligible} & ``Seats release not implemented\ldots!!!'' (367); the 689 TODO of another tool asks ``What about in cancel\_reservation?''. Never restored; per-episode reset (Reset Leak withdrawn). \\
\rowcolor{tblalt}
4 & MedAgentBench @ \texttt{9926011} & write graders, \texttt{refsol.py} (SHA-256-pinned; not in repo) & Ungrounded Oracle (evaluator property, \S\,\ref{sec:prelim}) & n/a (grader source) & ``POST request accepted'' gates \texttt{extract\_\brk posts}, graded from transcript, not live state. \\
5 & AgentDojo @ \texttt{089ed46} & \texttt{update\_\brk scheduled\_\brk transaction}, \texttt{banking\_\brk client.py:\brk 115-151} & Ignored Argument + Phantom Effect & agent-visible (\texttt{docstring}) & \texttt{recurring} guarded by truthiness (\texttt{True} only); returns ``\ldots updated'' unconditionally. \\
\rowcolor{tblalt}
6 & AgentDojo @ \texttt{089ed46} & \texttt{reserve\_\brk car\_\brk rental}, \texttt{travel\_\brk booking\_\brk client.py:\brk 382-400} & Ignored Argument & agent-visible (\texttt{docstring}) & \texttt{end\_time} dropped from state, kept in success string. No task exercises this tool. \\
7 & MM-ToolSandbox @ \texttt{1e8e932} & \texttt{venmo\_\brk social}, \texttt{mini\brk /venmo.py:\brk 464-470} & Ignored Argument & agent-visible (\texttt{docstring}) & \texttt{sort\_by} documented twice, never forwarded (listing branch; the observable is the forwarded call, where the two sibling arguments are present). \\
\rowcolor{tblalt}
8 & AgentDojo @ \texttt{089ed46} & \texttt{invite\_\brk user\_to\_\brk slack}, \texttt{slack.py:\brk 93-103} & Ignored Argument & agent-visible (\texttt{docstring}) & ``should be sent'' to \texttt{user\_\brk email}; body never reads it (static check). \\
\bottomrule
\end{tabular}

\end{table*}

The central chain is MedAgentBench's: Finding 4 makes Finding 1's omission unobservable from inside the benchmark, the grader admitting evidence only when it matches Finding 1's success string. MedAgentBench's paper reports ``Action SR'', the write-task success rate, for all 12 evaluated models in its Table 3 \cite{medagentbench25}, ranging 0.00\% to 71.33\%, and its abstract headlines 69.67\% overall SR for Claude 3.5 Sonnet v2, a weighted blend of Query SR and this Action SR. The file containing the POST branch is unchanged at our pinned commit, so every Action SR value was produced with no write occurring; its own \S\,2.4.1 calls these ``rule-based sanity checks to verify the correctness of the payload of POST requests,'' which is what the numbers measure, never whether any clinical record changed. We claim no causal link from any defect to any specific published number. The answer to Q1: each benchmark carries one headline-eligible defect class in the anchor audit, and in the clinical one a tool reports a write it never makes to a grader that takes the report as its evidence.\looseness=-1

\subsubsection{Q2, which verdicts are exposed, and whether any changes} Finding 8 is untraced in Table~\ref{tab:scoreatrisk}, since the email it drops has no state field for step 1 to name, and so are Q3's two later instances. tau2-bench's airline domain grades \texttt{cancel\_reservation} by hashing the full database between gold and predicted runs (\texttt{toolkit.py:242-244}), so all 50 airline tasks are exposed on basis \texttt{whole\_state\_hash}, a measure of the evaluator's coarseness, not of the defect. Telecom grades \texttt{refuel\_data} through per-task assertion functions, and the intersection is exact. For 1,135 of 2,285 tasks the verdict's derivation includes a field the defective precondition lets through unconditionally: 1,120 read the written attribute (\texttt{exact\_field}) and 15 the collection alone (\texttt{collection\_only}), summed only to give the union. The 2,285 is the full generated task file, not the roughly 114-task subset tau2-bench's paper evaluates; the pinned commit sits in the lineage behind its published leaderboard, and both audited defects are unpatched across that span \cite{tau2bench25}.\looseness=-1

\begin{table*}[t]
\caption{Score-at-risk by defect, basis and grounding. Exposed counts exposure only (\S\,\ref{sec:trace}): bases never pool, \texttt{unresolved} rows are forced in by the fallback, and no row establishes reach or a verdict change. Gold-call: tasks whose gold actions call the tool, which replay and positive controls use. Oracle grounding: oracles read stored state, the transcript, or both (mixed). n/c is the score-at-risk ledger's \texttt{not\_\brk computable\_\brk appworld\_\brk unreachable}.}
\label{tab:scoreatrisk}
\centering
\scriptsize
\begin{tabular}{@{}p{0.235\textwidth}p{0.13\textwidth}p{0.235\textwidth}>{\centering\arraybackslash}p{0.09\textwidth}>{\centering\arraybackslash}p{0.08\textwidth}p{0.18\textwidth}@{}}
\toprule
\rowcolor{tblhead}
\textbf{Benchmark : tool} & \textbf{Defect class} & \textbf{Basis} & \textbf{Exposed / total} & \textbf{Gold-call} & \textbf{Oracle grounding} \\
\midrule
tau2 airline : \texttt{cancel\_\brk reservation} & Partial Effect & \texttt{whole\_\brk state\_\brk hash} & 50 / 50 & 7 & state\_\brk grounded \\
\rowcolor{tblalt}
tau2 telecom : \texttt{refuel\_data} & Unenforced Precondition & \texttt{exact\_field} (1120) + \texttt{collection\_\brk only} (15) & 1135 / 2285 & 1120 & state\_\brk grounded \\
AgentDojo banking : \texttt{update\_\brk scheduled\_\brk transaction} & Ignored Argument & \texttt{exact\_field} (1) + \texttt{unresolved} (2) & 3 / 16 & 4 & state\_\brk grounded \\
\rowcolor{tblalt}
AgentDojo travel : \texttt{reserve\_\brk car\_\brk rental} & Ignored Argument & \texttt{exact\_field} (1) + \texttt{unresolved} (18) & 19 / 20 & 0 & mixed \\
MedAgentBench : post-write & Phantom Effect & transcript (not state) & undefined & n/a & 60 transcript / 90 mixed / 150 no-oracle / 0 state (of 300) \\
\rowcolor{tblalt}
MM-ToolSandbox : \texttt{venmo\_\brk social} & Ignored Argument & n/c & n/c & n/c & n/c \\
\bottomrule
\end{tabular}
\end{table*}

MedAgentBench is classified per grader function, since a names-only rule would call nearly all 300 cases exposed: the grader-checked field and the field the tool should have written share a name. Of 300 cases, 150 are query tasks with no write component; of the 150 action tasks, 60 are graded unconditionally from the transcript and 90 gate on live state but still grade the write from the transcript, and zero read the write back from FHIR state. No action verdict's derivation includes the record the write should have produced; its source record for the write is the transcript. Exposure is therefore undefined for the 60 and indeterminate for the 90, never zero: ``oracle not state-grounded.'' MM-ToolSandbox's row is not computable rather than zero: step 1 names the \texttt{sort\_by} field Finding 7 drops, and steps 2 and 3 stop at the \texttt{appworld} package.\looseness=-1

Exposure is a field-read rule and the gold-call set a gold-invocation rule, so the two need not nest. For Finding 5 they share no resolved row. Only \texttt{recurring} admits a legitimate falsy value, so it alone carries exposure, while the four gold-call tasks vary \texttt{amount} or \texttt{recipient}. The fallback forces in, at confidence \texttt{unresolved}, the rows the extractor cannot read: banking \texttt{UserTask9} and \texttt{UserTask10} (whole-environment equality), the first also a gold-call task, and 18 of 20 travel tasks (\texttt{TravelDeepDiff}). Neither AgentDojo count is therefore a bound. No shipped travel task calls \texttt{reserve\_car\_rental}, so the 19 of 20 is exposure with no shipped task reaching the defect, and nothing beyond exposure is claimed.\looseness=-1

\paragraph{Gold-trajectory replay} Reach on shipped paths. Findings 2 and 3 (F2, F3) carry mechanically verified one-hunk patches: \texttt{seats\_after} is 0 unpatched and 3 patched, and a suspended-line refuel succeeds unpatched and raises \texttt{ValueError} patched. Ten trajectories were replayed from each finding's own gold reference solution, five per finding by seeded sample, zero discarded; agent trajectories were unavailable offline, so gold references were substituted, a disclosed deviation from the plan. All ten share one outcome, exposed and not reaching the defect: zero meet either trigger condition, so zero verdicts change. For F3 no verdict change was possible by construction, gold and agent runs executing the same still-defective tool; for F2 one was possible and did not occur, because no gold trajectory calls the tool on an inactive line and all 1,120 telecom tasks whose gold actions call it issue the same one.\looseness=-1

\paragraph{Positive controls} Reach and verdict change on constructed paths. The null cannot separate a defect no path reaches from one no gold path happens to reach. A second experiment, planned after the null and so post hoc, though registered before it ran, constructs the path by hand. For F2, each of the 1,120 gold action sequences is kept unchanged with \texttt{suspend\_line} inserted before the gold \texttt{refuel\_data} call (C1) or with \texttt{resume\_line} also inserted after it (C2); for F3, each of the 7 airline gold-call tasks gets a \texttt{book\_reservation} and a \texttt{cancel\_reservation} of that booking appended. The F2 gold-call set is the same 1,120 tasks Table~\ref{tab:scoreatrisk} tags \texttt{exact\_field}, graded on environment assertions alone, so those exposed tasks are the scored tasks; they collapse to 1,104 distinct action sequences (largest group 2), but every task ran independently, its gating assertions being unique. Each (task, construction) ran against the unpatched and the patched tool, 4,494 rows, no model calls, seed 1 (Table~\ref{tab:poscontrol}). The independent verdict is a predicate we wrote after seeing the defect, reported beside the shipped verdict, never in its place.\looseness=-1

\begin{table}[t]
\caption{Positive controls per construction and tool arm.}
\label{tab:poscontrol}
\centering
\scriptsize
\setlength{\tabcolsep}{2.5pt}
\begin{tabular}{@{}lrrrrrrrr@{}}
\toprule
\rowcolor{tblhead}
 & & \multicolumn{2}{c}{\textbf{Shipped PASS}} & \multicolumn{2}{c}{\textbf{Indep. PASS}} & \multicolumn{2}{c}{\textbf{State changed}} & \makecell{\textbf{Verdict}\\\textbf{changed}} \\
\cmidrule(lr){3-4}\cmidrule(lr){5-6}\cmidrule(lr){7-8}
\rowcolor{tblhead}
\textbf{Constr.} & \textbf{Tasks} & \textbf{unp.} & \textbf{pat.} & \textbf{unp.} & \textbf{pat.} & \textbf{unp.} & \textbf{pat.} & \\
\midrule
F2 C1 & 1120 & 312 & 0 & 0 & 1120 & 1120 & 0 & 312 \\
\rowcolor{tblalt}
F2 C2 & 1120 & 1120 & 0 & 0 & 1120 & 1120 & 0 & 1120 \\
\midrule
F3 & 6 of 7 & 0 & 0 & 0 & 6 & 6 & 0 & 0 \\
\midrule
\multicolumn{9}{@{}p{\columnwidth}@{}}{unp.: unpatched; pat.: patched; the independent verdict is unanimous in every arm, so each 2$\times$2 is its margins. F3's seventh task was infeasible (\texttt{Too many reservations}).} \\
\bottomrule
\end{tabular}
\end{table}

Under C2 the shipped evaluator awards full reward to all 1,120 unpatched trajectories and fails all 1,120 once the one-hunk repair is applied, state diff and independent verdict moving with the patch. It was rewarding a refuel that bills a line the docstring says must be Active, the pre-registered reading. The 1,120 is fixed by construction, one path per gold-call task: on such a path the evaluator cannot distinguish the defect from correct behaviour, which says nothing of how many tasks are affected in practice. Under C1 the shipped verdict changes on 312 tasks and the other 808 fail unpatched as well, because suspending the line trips user-side assertions unrelated to the refuel, the confound C2 was pre-registered to remove. F3 is the plan's second outcome: airline grades by whole-database hash, so a constructed booking already differs from gold, both arms fail for a reason unrelated to the defect, and zero verdicts change while the independent predicate separates them. Table~\ref{tab:scoreatrisk}'s 50 airline rows remain exposure, with reach and no verdict change established on the six constructed paths. The gold-trajectory null stands as the pre-registered result, and neither experiment says how often an agent takes such a path: F3's replays cover five of seven gold-call tasks, the other two, read by hand, cancel nothing booked in-episode, F2's one identical call, and the constructed paths are ours. The answer to Q2: exposure is wide, reach on shipped gold paths is nil, and on every path built to isolate F2 the shipped evaluator rewards the defect and fails its repair, an existence result, not a frequency.\looseness=-1

\subsubsection{Q3, what the checker adds beyond reading source}   The comparison is against reading source, which is how the findings were made: seven of the eight instances of Table~\ref{tab:findings} surfaced by manual audit, Finding 8 by the static half, none by the dynamic checker, whose role on them, where it runs, is to confirm against executed state and trace.\looseness=-1

\paragraph{Static scanner} The scanner is the checker's static half (\S\,\ref{sec:checker}), so Q3 compares the two halves with manual reading, not with an outside method; several checks took the shape of findings in hand, so its hits on them are partly by construction.  It ran over the 49-tool scanner population (\S\,\ref{sec:setup}) under a rule fixed before it ran: a flag is a true positive only if it names both the tool and the confirmed site, every other flag, none discarded as untriaged, a false positive. It raised 23 flags, 14 true and 9 false positives (2 at the wrong site of \texttt{venmo\_social}, 7 on tools with no finding), and flagged 14 of the 17 confirmed sites: it misses Finding 5's success return and both blind calendar sites, raising no flag anywhere in \texttt{calendar\_client.py}. Over whole files a maintainer would see 32 flags (14 true, 18 false), the nine extra all false and on uncontracted neighbours; the population figure is the headline because the plan fixed the population, a choice that runs against the dynamic half.\looseness=-1

\paragraph{Blind audit} The twelve sit in six modules, four never before yielding a finding; \texttt{add\_calendar\_\brk event\_participants} was declared a prior candidate. Of the other eleven, 8 CONFORM, 1 VIOLATES and 2 are UNTESTABLE; UNTESTABLE is 14 of 65 clause rows. The VIOLATES is \texttt{cancel\_calendar\_event}, Partial Effect at the agent-visible tier. Its docstring says the event ``will be marked as canceled and no longer appear in the calendar,'' but \texttt{Calendar.cancel\_event} (\texttt{calendar\_\brk client.py:58-61}) only sets \texttt{status}, which nothing in the file reads. \texttt{get\_by\_day} filters on date and \texttt{search\_events} on text alone, so both listing tools still return a cancelled event. Confirmed by source reading, it is the first defect the dynamic checker has found in code nobody had flagged in advance. Reported beside that result, never in it, \texttt{add\_\brk calendar\_\brk event\_\brk participants} is source-confirmed only: its docstring promises the new participants an email, but unlike its three sibling calendar tools it takes no \texttt{inbox} dependency, so no code path can send one. It, \texttt{create\_calendar\_event} and \texttt{send\_email}, 3 of the twelve, are UNTESTABLE on one crash, \texttt{TypeError: unhashable type: 'list'}, from the probe-deduplication hash on a list-valued argument no earlier contract had.\looseness=-1

\paragraph{Held-out audit} The six held-out tools ran first. Prior knowledge was declared: \texttt{update\_user\_info} was already a ledger candidate, and the scanner had run repository-wide. Five CONFORM; \texttt{update\_user\_info} VIOLATES on four agent-visible effect clauses, repeating Finding 5's truthiness-guard shape (\texttt{user\_account.py:61-68}), a second instance of that class, not a new cell. UNTESTABLE is 3 of 43 clause rows.\looseness=-1

\paragraph{The complete AgentDojo surface}   The setup's implementation-side rule enumerated 25 mutating tools across AgentDojo's v1 suites, and the artifact holds exactly those 25 contracts, complete under the rule, though the rule admits no tool that advertises a change and writes nothing, MedAgentBench's shape. Detection is not complete. Five of the 25 carry a checker-confirmed VIOLATES, each confirmed by source reading: Findings 5, 6 and 8, the held-out \texttt{update\_user\_info} and the blind \texttt{cancel\_calendar\_event}. Three are UNTESTABLE on the crash above, the source-confirmed \texttt{add\_\brk calendar\_\brk event\_\brk participants} among them. The remaining 17 CONFORM under a procedure whose recall is low (Q4), which does not establish absence.\looseness=-1

The rate we report is therefore at least 5 of 25, UNTESTABLE kept in the denominator; over the 22 tools the checker scored it is 5 of 22, higher only because three drop out for a checker limitation. The procedure detects exactly these five, so the rate is a lower bound on the tools that diverge from the advertised surface as our contracts read it. It is per tool, a tool possibly carrying several instances, and is not a prevalence estimate, bug rate or severity claim, nor about the other three benchmarks. The answer to Q3: beyond the anchor findings, which the scanner nearly matches, the checker adds one confirmed defect in code nobody had flagged, and for one benchmark a complete though not uniformly blind surface on which at least 5 of 25 tools diverge, four known in advance.\looseness=-1

\subsubsection{Q4, how reliable the checker is}   In mutation testing's vocabulary \cite{jiaharman11}, synthetic faults (mutants) go into the real tools and into a small toy domain shipped in the artifact, and surviving mutants bound the misses from below. Both arms evaluate the whole pipeline, contract, probes and checker, and the plan fixed before any mutant existed what each can license: the closed-world arm, mutations of the six classes, ``the checker detects violations of the clauses we wrote,'' never ``tool-contract defects''; the open-world arm, mutations made without regard to the taxonomy, a lower bound on what the pipeline misses, decomposed by cause. Every interval bound is the lower limit of a two-sided 95\% Wilson interval; recall rates are design-effect-adjusted for clustering by tool (Kish design effect, Rao-Scott one-way ANOVA ICC estimator), the effective denominator beside each; pooled precision and escape rates are unclustered at their $n$.\looseness=-1

\begin{table}[t]
\caption{Real-tools closed-world recall by mutation operator; eff.~n (design-effect-adjusted) feeds every Wilson bound.}
\label{tab:recall}
\centering
\scriptsize
\setlength{\tabcolsep}{2pt}
\begin{tabular}{@{}lccccccc@{}}
\toprule
\rowcolor{tblhead}
\textbf{Operator} & \textbf{n drawn} & \textbf{eff.~n} & \textbf{detected} & \textbf{missed} & \textbf{untestable} & \textbf{error} & \textbf{Recall (Wilson LB)} \\
\midrule
M-PHANTOM & 7 & 7 & 5 & 0 & 0 & 2 & $\geq$0.359 \\
\rowcolor{tblalt}
M-PRECOND & 7 & 5 & 1 & 6 & 0 & 0 & $\geq$0.020 \\
M-IGNARG & 21 & 15.13 & 4 & 12 & 1 & 4 & $\geq$0.066 \\
\rowcolor{tblalt}
M-PARTIAL & 8 & 6 & 1 & 7 & 0 & 0 & $\geq$0.018 \\
M-INVAR & 8 & 8 & 0 & 8 & 0 & 0 & $\geq$0.000 \\
\rowcolor{tblalt}
M-RESET & n/a & n/a & n/a & n/a & n/a & n/a & no data \\
\bottomrule
\end{tabular}
\end{table}

\paragraph{Closed-world recall} Recall is low throughout (Table~\ref{tab:recall}). A post hoc diagnostic traces each of the 33 real-tool misses to its cause. For 2 no clause covers the mutation. For 29 a clause exists but no probe of an independent frozen corpus revealed any behavioural difference, the mutant unreached or inert. The remaining 2 are a probe gap and a target not found. Only the first bears on contract coverage; the 29 point to the probe stage, part of the pipeline, or to an inert mutant. M-INVAR's 0.000 therefore reads as ``no probe ever reached a mutated invariant path,'' not ``the checker cannot detect invariant breaks.'' M-PHANTOM depends on a biconditional success signal, declared on 18 of 19 tau2-bench, 1 of 7 AgentDojo and 0 of 5 MM-ToolSandbox contracts. Only M-PRECOND and M-IGNARG met the pre-registered $\sim$25-per-class target, counting real and toy sites together, which is why the real-tool-only n-drawn column reads 7 and 21; the other four fall short because the corpus is structurally small, not by choice.\looseness=-1

\paragraph{Precision} The checker raised zero false positives against any injected mutant in the closed-world arm, across a pooled toy-plus-real sample of 25 flags (TP=25, FP=0; 24 semantics-preserving controls scored alongside). Precision is $\geq$0.867, never reported without the negative controls below.

\paragraph{Negative controls} Five inputs built to resemble defects without being defects ran against the unmodified checker, each verdict predicted before the run, none added or removed afterwards. Two flagged: \texttt{update\_\brk reservation\_\brk baggages} called with the reservation's current baggage counts (control 1), and \texttt{reserve\_\brk hotel} given an ISO date with and without a seconds field (4b). Both are false positives of the Ignored Argument rule, by the mechanism the plan named, and its repair is unapplied (\S\,\ref{sec:threats}): whole-post-state equality cannot tell an effective argument whose effect is identical across the values tried from one the code drops. The predictions were wrong on 3 of 5, each traced to source. Control 2 CONFORMS, not flagged: \texttt{add\_user\_\brk to\_channel} appends an existing member with no deduplication guard. Control 4a is UNTESTABLE, not flagged: Python equality collapses \texttt{gb\_amount} 2 and 2.0 in the distinctness gate. Control 3, the disclosed flight-inventory shortcut of \texttt{update\_\brk reservation\_\brk flights} (line 689), is UNTESTABLE rather than CONFORMS: no synthesizer exists for a \texttt{List[\brk FlightInfo]} argument, so the probe crashes the tool before its effect code.\looseness=-1

\paragraph{Open-world escape rates} On the toy domain, the Cosmic Ray mutation tool scored 368 mutations across 4,048 (mutation, tool) pairs; 229 were behaviourally live, 225 of them escaped the checker, and the missed share is a Wilson lower bound of 0.956 (point estimate 0.983). By cause, 119 of the 225 escapes had no clause covering the mutation, 86 had a clause whose probes never drove the tool into the revealing state, and 20 crashed the checker before evaluation; only the 119 bear on contract coverage. Only the mechanical declared-volatile-field rule ran, so 0.956 bounds the escape rate under that one rule, overstates the pipeline's blindness if anything, and is not evidence against the closed-world numbers. On tau2-bench telecom, 465 mutations across 2,325 pairs against the domain's five tools with no finding or candidate left zero behaviourally live, since the probe stage's recorded-real-call component does not exist and its placeholders fail tau2's entity-keyed checks. That zero denominator is no evidence of a well-covered contract. The answer to Q4: the checker is precise on what it flags, zero false positives on 25 injected flags against two among five negative controls, and misses most of what it should, mostly because no probe reveals a mutant a clause already covers.\looseness=-1

\subsubsection{Q5, whether contract authoring is stable}\label{sec:q5} Two annotator sets, both agent sessions on one model family and only B blind, wrote contracts for the six tools of Findings 2, 3 and 5--8 against one pre-committed probe corpus; this measures whether a sentence becomes the same predicate twice, not whether two people would agree. \texttt{cancel\_reservation}, of which B had read a worked example, is a contaminated stratum never pooled with the blind result. Four of the five blind tools contribute pairs; \texttt{venmo\_social} contributes none, a segmentation artifact that also cost it the clause behind Finding 7. Under the protocol's rule, which counts UNTESTABLE against anything as a disagreement, the blind stratum agrees on 98 of 99 probe comparisons; the one disagreement, \texttt{reserve\_car\_rental}'s \texttt{arg.end\_time} (VIOLATES under A, UNTESTABLE under B), is a missing ISO-string annotation, not a semantic split. A post hoc comparability filter restricted to successful calls, which landed with the result rather than the protocol, gives 98 of 98 across 16 clause pairs by excluding it. The answer to Q5: on the blind stratum, 98 times in 99.\looseness=-1

\subsection{Threats to Validity}
\label{sec:threats}

\textbf{Taxonomy, grammar and corpus limits.} Six classes are not the space of possible defects. The frame-path grammar cannot exclude a single key, so ``no other reservation is modified'' collides with the advertised effect, and the shipped \texttt{cancel\_reservation} contract states a weaker claim on an unrelated collection. The probe stage's missing recorded-real-calls component could move the conclusions most: it leaves the real-benchmark arm no usable escape rate and may account for the 29 of 33 closed-world misses no probe revealed. M-RESET has no data, not a zero recall: reset lives in the adapter layer, which rebuilds environments wholesale, and no tool file in the four benchmarks defines a state-restoring function.\looseness=-1

\textbf{Frozen checker, pre-registration, and known limits.} Every number is scored under one checker freeze, left unpatched once scoring began. The whole-post-state rule's two confirmed false positives, the equality gate that collapses 2 and 2.0, and the list-argument gaps that made three blind tools UNTESTABLE therefore stand unrepaired; the first's pre-committed repair would need a third freeze. The two flagged controls make the mutation-arm zero no precision claim about shipped tools; the Ignored Argument findings rest on source reading, not the rule alone; the same gates bound the recall table. One earlier refreeze is disclosed: a sweep against the first freeze found four taxonomy claims with no code path behind them (Reset Leak and Invariant Break verdicts, 35 frame clauses, and 17 of 19 tau2 contracts were silently never dynamically exercised), all fixed before the second freeze scored any number here. The subprocess boundary absorbs exceptions the checker never sees: in the closed-world run a toy adapter's \texttt{invoke()} caught only its declared error type, and a raw exception past a deleted guard aborted the mutant with no violation. It sits under every real adapter; whether an absorbed crash scores as detected or silent is unchecked and bounds every real-tools recall number. \texttt{git init} ran on 2026-08-21, after most design work, and the first commit imported the project as one tree, so we do not claim any pre-registration preceded the decisions it was meant to freeze; later commits verify that the first freeze predates the first scored mutant, the agent-experiment plan the first recorded trajectory, and the round-two plans every round-two number.\looseness=-1

\textbf{The trace's second reader is a model session.} One author wrote the read-site extractor and judged what it ought to find. The pre-registered second reading so far comes only from the plan's fallback, a model session given the 40 evaluator functions and instructions alone, blind to the extractor, findings and paper: weaker evidence than a human reader. It named fields for 22 functions; on all 22 the extractor's set was a strict superset, none missed, at a median 1.7$\times$ the annotator's count, worst 10$\times$. On the other 18, whole-environment equalities or functions reading no state, it named none, and the extractor had independently marked 14 \texttt{unresolved}. Six travel functions were read partially, their gating helpers absent from the package, since shipped. Every disagreement is listed in the artifact, with no coefficient; Table~\ref{tab:scoreatrisk}'s 20 \texttt{unresolved} rows remain the fallback's output, not a measurement.\looseness=-1

\textbf{Anchor-driven benchmark selection.} Benchmarks entered on known defects or to test generalization (\S\,\ref{sec:setup}), so no prevalence claim across them is supportable. Within AgentDojo the findings are no longer a sample, but the blind audit's one defect in eleven is the only evidence that the dynamic checker finds anything beyond code already suspected, and elsewhere every finding is anchor-driven.\looseness=-1

\section{Conclusion}
\label{sec:conclusion}

This paper tested the tool layer beneath the surveyed taxonomies' categories in four benchmarks, and so the provenance of their scores. For AgentDojo every enumerated mutating tool has a contract and at least 5 of the 25 diverge from their advertised surface, a rate for that benchmark alone. The anchor-driven audits are existence proofs, not rates; an exposure count is a due-diligence flag, not a correction to a published score; and the checker's low recall means a CONFORM verdict does not establish conformance. Maintainers can ground write graders in stored state, not the tool's success string, audit the tool layer, not only tasks and gold solutions, and disclose every deliberate simplification where an agent reads it.\looseness=-1

\paragraph{Coordinated disclosure} Three teams received findings, reproductions and repairs on 2026-09-12 (MedAgentBench issue 10, tau2-bench issue 541, AgentDojo issue 194), each headline finding re-checked that day on its default branch. MM-ToolSandbox disables issues and declines external reports in its \texttt{SECURITY.md}, so Finding 7 has no channel. The held-out and blind findings, with \texttt{add\_\brk calendar\_\brk event\_\brk participants} labelled source-read only, were filed on the AgentDojo thread on 2026-09-24. By a rule fixed in advance, the disclosure log reports responses in substance, records non-response by date with no inference from silence, and marks contested findings with the maintainer's reasoning.\looseness=-1

\paragraph{Acknowledgment} AI tools assisted with the analysis code and experiment runs; the authors reviewed all content and are responsible for it.

\balance
\bibliographystyle{IEEEtran}
{\renewcommand{\footnotesize}{\scriptsize}\def\IEEEbibitemsep{0pt}\bibliography{references}}

\end{document}